\documentstyle[multicol,pra,aps,amsfonts]{revtex}

\newcommand{\beq}{\begin{equation}}
\newcommand{\eeq}{\end{equation}}
\newcommand{\beqy}{\begin{eqnarray}}
\newcommand{\eeqy}{\end{eqnarray}}

\newenvironment{Definition*}{{\bf Definition}}{}

\makeatletter
\def\@beginTheorem#1#2{\trivlist \item[\hskip \labelsep{\bf #1\ #2}]}
\def\@opargbegintheorem#1#2#3{ \trivlist
      \item[\hskip \labelsep{\bf #1\ #2\ (#3)}]}
\makeatother
\makeatletter
\def\@beginLemma#1#2{\trivlist \item[\hskip \labelsep{\bf #1\ #2}]}
\def\@opargbeginLemma#1#2#3{ \trivlist
      \item[\hski
 Hence we have the same statements
about the increase of the supports for increasing depth, where
the local transformations are not counted for the depth.
p \labelsep{\bf #1\ #2\ (#3)}]}
\makeatother

\makeatletter
\def\@beginDefinition#1#2{\trivlist \item[\hskip \labelsep{\bf #1\ #2}]}
\def\@opargbeginDefinition#1#2#3{ \trivlist
      \item[\hskip \labelsep{\bf #1\ #2\ (#3)}]}
\makeatother

\makeatletter
\def\@beginCorollary#1#2{\trivlist \item[\hskip \labelsep{\bf #1\ #2}]}
\def\@opargbeginCorollary#1#2#3{ \trivlist
      \item[\hskip \labelsep{\bf #1\ #2\ (#3)}]}
\makeatother

\makeatletter
\def\@beginExample#1#2{\trivlist \item[\hskip \labelsep{\bf #1\ #2}]}
\def\@opargbeginExample#1#2#3{ \trivlist
      \item[\hskip \labelsep{\bf #1\ #2\ (#3)}]}
\makeatother

\def\C{{\mathbb{C}}}

\title{Complexity of decoupling and  time-reversal
for $n$ spins
 with pair-interactions: \\ Arrow of time in quantum control}

\author{Dominik Janzing\thanks{Electronic address: janzing@ira.uka.de},
Pawel Wocjan, and Thomas Beth} \address{Institut f\"ur Algorithmen und
Kognitive Systeme, Am Fasanengarten 5, D--76\,131 Karlsruhe, Germany}

\begin{document}

\maketitle

\begin{abstract}
Well-known Nuclear Magnetic Resonance experiments show that the time
evolution according to (truncated) 
dipole-dipole interactions between $n$ spins
can be inverted by simple pulse sequences. Independent of $n$, the
reversed evolution is only two times slower than the original one.
Here we consider more general spin-spin couplings with long range. We prove
that some are considerably more complex to invert since the number of
required time steps and the slow-down of the reversed evolutions are
necessarily of the order $n$. Furthermore, the spins have to be
addressed separately. We show for which values of the coupling
parameters the phase transition between simple and complex
time-reversal schemes occurs.
\end{abstract}

\begin{multicols}{2}
Due to wide applications of Nuclear Magnetic Resonance in medicine,
chemistry, and physics much effort has been undertaken to understand
the dynamics of nuclear spins in solids and liquids
\cite{Maciel,VlBoe}. Nowadays, NMR plays an important role in the
first experimental realizations of quantum computation \cite{nielsen}.

Phenomenologically, the dynamical evolution of a single spin seems to
be a relatively simple relaxation and dephasing process of an open
quantum system, formally described by the Bloch equations
\cite{ernst,slichter}. However, the fact that not all
dephasing and relaxation processes are irreversible is important for
practical purposes since refocusing techniques are applied with great
success \cite{ernst,haeberlen,slichter,rhim}.

The simplest version of refocusing is possible when dephasing of the
(uncoupled) spins is caused by spatial inhomogeneities of the strength
of the magnetic field. In this case the time evolution can be reversed
by sandwiching the natural dynamics by $180^\circ$--rotations of all
spins around an axis orthogonal to the magnetic field. More
sophisticated versions of refocusing are necessary if dephasing and
relaxation processes are caused by spin-spin interactions. An
important example is the so-called dipole-dipole coupling \cite{rhim}.

We describe this interaction formally as follows. Let $(\C^2)^{\otimes
n}$ be the Hilbert space of $n$ spins and $\sigma_\alpha^{(k)}$ be the
Pauli matrix $\sigma_\alpha$ acting on the $k$-th spin for $1\leq
k\leq n$ and $\alpha=x,y,z$. Then the so-called truncated
dipole-dipole Hamiltonian $H_d$ is given by
\[
H_d:=\sum_{k<l} w_{kl}\, \big( \sum_\alpha \sigma_\alpha^{(k)}
\sigma_\alpha^{(l)} - 3 \sigma_z^{(k)} \sigma_z^{(l)} \big) \,,
\]
where $w_{kl}$ is the strength of the interaction between the spins
$k$ and $l$. 

We assume that all local unitary operations on the spins can be
produced in arbitrarily small time (on a time scale governed by the
strength of the coupling). This is often called the {\em fast control
limit}. In practice this is done by external electro-magnetic pulses.

Consider a pulse implementing a rotation of each spin around the
$y$-axis by $90^\circ$, i.\,e., if the spin is in $z$-direction it is in
$x$-direction after the rotation. Formally, one applies a unitary
transformation on $(\C^2)^{\otimes n}$ of the form
\[
v_y:=u_y\otimes u_y \otimes \cdots \otimes u_y
\]
with $u_y^\dagger\sigma_z u_y = \sigma_x$. Now we focus on the effect
of the following scheme: (1) implement $v_y$ by the corresponding
pulse, (2) wait the time $t$, and (3) implement $v_y^\dagger$, i.\,e., the
inverse of $v_y$. The net effect of the scheme is an evolution
\[
v_y^\dagger \exp(-i H_d t) v_y = \exp(-i v_y^\dagger H_d v_y \,t)\,,
\]
i.\,e., a dynamics according to the Hamiltonian $v_y^\dagger H_d v_y$.
We choose a second rotation $u_x$ around the $x$-axis such that
$u_x^\dagger \sigma_z u_x=\sigma_y$ and set $v_x:= u_x\otimes
u_x\otimes\cdots\otimes u_x$.  It is decisive to note that
\[
-H_d= v_y^\dagger H_d v_y + v_x^\dagger H_d v_x\,.
\]
This has the interesting consequence that for small times $\epsilon$
one has
\[
v_x^\dagger \exp(-i H_d \epsilon) v_x\, v_y^\dagger \exp(-iH_d\epsilon) v_y
\approx \exp( +i H_d \epsilon)
\]
up to terms of second order of $\epsilon$. This first order
approximation is usually referred to as {\em average Hamiltonian
theory} \cite{ernst}. The intuitive meaning is that the time evolution
according to $H_d$ can be approximatively inverted by dividing the
time $t$ into small intervals of length $\epsilon$ and interspersing
the natural time evolution by the above pulses.  The inverted dynamics
is two times slower than the natural one. We express it by saying that
the {\it time overhead} of inverting the Hamiltonian $H_d$ is $2$. The
task of time-reversal is a special case of the general problem of
simulating Hamiltonians that has recently been considered by several
authors
\cite{knill,leung,stoll,dodd,graph,bett,reval,finite,NN,eff,vid,homo}.

Note that the decoupling and inversion schemes for the dipole-dipole
Hamiltonian is simple with respect to three aspects: (a) it does not
use selective pulses, i.e., in each time step the same unitary
transformation is applied to all spins. (b) The number of steps of
this inversion scheme does not increase with $n$ since it is always
$2$, and (3) the time overhead does not increase with $n$ since it is
also $2$.

The purpose of this article is to show that there are interactions
that are considerably more complex to invert with respect to all three
criteria.  The most general spin-spin interaction on $n$ spins is
given by
\[
H_J:=\sum_{k<l} \sum_{\alpha,\beta} 
J_{kl;\alpha\beta} \sigma_\alpha^{(k)} \sigma_\beta^{(l)}\,,
\]
where $J$ is chosen to be a real symmetric $3n \times 3n$-matrix with
zeros for $k=l$. Note that the symmetry of the coupling matrix $J$
does not imply any physical symmetry of the interaction. It is a
consequence of our redundant notation that turns out to be very
useful. The coupling matrix $J$ consists of $3\times 3$-blocks. The
$3\times 3$-matrix $J_{kl}$ given by the block at position $(k,l)$
describes the coupling between the spins $k$ and $l$. We have
$J_{lk}=J_{kl}^T$, i.\,e. the matrix describing the coupling between
the spins $l$ and $k$ is just the transpose of the matrix describing
the coupling between the spins $k$ and $l$. The blocks on the diagonal
are zero matrices.

Let ${\mathcal C}:=SU(2)\otimes SU(2)\otimes\cdots\otimes SU(2)$ be
the control group of local operations on the spins. As already
mentioned all operations in ${\mathcal C}$ are assumed to be implemented
arbitrarily
fast. The most general refocusing scheme (also called 
{\it time-reversal} or {\it inversion}) based on the first order
approximation above is given by rotations $v_1,v_2,\dots,v_N \in
{\mathcal C}$, and (relative) times $t_1,t_2,\dots,t_N$ such that the
average Hamiltonian
\[
\bar{H}:= \sum_j t_j v_j^\dagger H_J v_j
\]
is equal to $-H_J$. Here $N$ is the {\em number of time steps} and
$\tau:=\sum_{j=1}^N t_j$ is the {\em time overhead}. The time overhead
gives the slow-down of the inverted Hamiltonian. In the following we
consider both the number of time steps and the time overhead as
complexity measures. In contrast, if the sequences are chosen in such a way
that the average Hamiltonian is zero one has a {\it decoupling} scheme.
Each inversion scheme with $N$ steps 
defines a decoupling scheme with $N+1$ steps in a straightforward way
(note that the notion of time overhead does not make sense for decoupling).
Conversely, each
decoupling scheme can be converted to an inversion scheme as follows.
The equation 
\[
\sum_{j=0}^N t_j v_j^\dagger H_J v_j=0
\]
 implies
\[ 
\sum_{j=1}^N (t_j/t_0)  (v_j v_0^\dagger)^\dagger 
H_J (v_j v_0^\dagger)^\dagger =-H_J
\]
 by elementary
calculation. Therefore we shall restrict our attention to inversion.

The condition on inversion schemes can be rewritten by expressing the
action of the control operations on the coupling matrix $J$ as
follows. Note that any unitary operation $u\in SU(2)$ corresponds to a
rotation on the $3$-dimensional Bloch sphere via the relation
\[
u^\dagger \big( \sum_\alpha c_\alpha \sigma_\alpha \big) u = 
\sum_\alpha \tilde{c}_\alpha \sigma_\alpha\,,
\]
where the vector $\tilde{c}=(\tilde{c}_x,\tilde{c}_y,\tilde{c}_z)$ is
obtained by applying a rotation $U \in SO(3)$ on the vector
$c=(c_x,c_y,c_z)$. It is straightforward to verify that conjugation of
$H_J$ by $v:=u^{(1)}\otimes u^{(2)}\otimes\cdots\otimes u^{(n)}$ 
corresponds to
conjugation of $J$ by a block diagonal matrix of the form
\[
V:=U^{(1)}\oplus U^{(2)}\oplus\cdots\oplus U^{(n)}\in 
\bigoplus_{k=1}^n SO(3)\,.
\]
The condition for time inversion is hence given by
\begin{equation}\label{condition}
-J= \sum_j t_j V_j J V_j^T \,,
\end{equation}
where the orthogonal matrix $V_j$ corresponds to the unitary transformation
$v_j$ for $j=1,\ldots,N$. 
Time reversal and decoupling schemes that apply 
for general coupling $J$ have been presented in
\cite{stoll,leung}. These schemes can even be used when the coupling
$J$ is unknown. Here we focus on optimality criteria for schemes
referring to specific couplings and show that there are interactions
that cannot be inverted significantly more efficiently 
than an unknown interaction. 

We restrict our attention to interactions with an additional
symmetry, namely a Hamiltonian of the following form
\[
H:=\sum_{k<l} w_{kl} 
\sum_{\alpha,\beta} A_{\alpha\beta} \sigma_\alpha^{(k)}\sigma_\beta^{(l)}
\,.
\]
The matrix $W:=(w_{kl})$ is a real symmetric $n\times n$-matrix with
zeros on the diagonal. It describes the coupling strengths and the signs
 of the 
interactions between all spins. The matrix $A=(A_{\alpha\beta})$ is a real
symmetric $3\times 3$-matrix characterizing the type of the coupling. This means
that all spins interact with each other via the same interaction and that
only the coupling strength and the sign 
varies. It is important that in this special case
the coupling matrix $J$ can be expressed as a tensor product of $W$ and 
$A$, i.\,e. $J = W \otimes A$.

The matrix $W$ characterizes the coupling topology of the spins. In
the language of graph theory the matrix $W$ is the adjacency matrix of
a weighted graph. Therefore many results on graph spectra (eigenvalues
of the adjacency matrices) can be used to derive lower and upper
bounds on the number of time steps and the time overhead.

To discuss the complexity aspects we distinguish between the following
three cases:
\begin{enumerate}
\item 
$A$ is traceless.

All spins can be subjected to the same transformations in each time
step, the number of time steps and the time overhead are at most
$2$. 

\item
$A$ has negative and positive eigenvalues but $tr(A)\neq 0$.

The spins have to be addressed separately, the number of time steps
necessarily grows for increasing $n$. But the time overhead does not
depend on $n$. It depends only on the eigenvalues of $A$.

\item
$A$ is either positive or negative semidefinite, i.\,e., the non-zero
eigenvalues have the same sign.

Then the spins have to be addressed separately, the number of time
steps is at least $n-1$, and the time overhead is also at least $n-1$.

\end{enumerate}
To prove these statements we assume w.l.o.g. that the interaction
between all pairs is of the form
\[
a_x\sigma_x \otimes\sigma_x +a_y \sigma_y \otimes\sigma_y +a_z\sigma_z
\otimes \sigma_z\,,
\]
where $a_x,a_y,a_z$ are the eigenvalues of $A$. This can always be
achieved by rotating the reference frame \cite{graph}.

\vspace{0.3cm} 
{\bf Case 1.} 
Assume $A$ to be traceless. Let $S$ be a
rotation in the Bloch sphere that realizes the cyclic permutation of
the axis according to $x\rightarrow y \rightarrow z \rightarrow
x$. Then we have
\[
\sum_{j=0}^2 V^j J {V^{j}}^T =\mathbf{0}\,,
\]
where $V:=S\oplus S\oplus\cdots\oplus S$ and  $V^j$ is the $j$th power of 
$V$. Here $S$ is a rotation in
the Bloch sphere that realizes the cyclic permutation of the axis
according to $x\rightarrow y \rightarrow z \rightarrow x$. The $j$th power
of $V$ is denoted by $V^j$. Hence the Hamiltonian is inverted by a
sequence of length $2$. Due to the equation
\[
-J = V J V^T + V^2 J {V^2}^T
\]
the time overhead of this inversion scheme is $2$.

\vspace{0.3cm}
{\bf Case 2.} 
The fact that the spins have to be addressed separately has 
also been noted in \cite{homo}. It  can 
be seen as follows. If all spins are subjected to the same transformation
each $3\times 3$--block of $J$ is conjugated by an element of $SO(3)$.
Such a conjugation preserves the trace of each block matrix. This trace is
even preserved under positive linear 
combination, i.e. the trace of each resulting $3\times 3$ matrix
has the same sign as the trace of the original matrix.
Therefore a sequence applying the same transformation to all spins
can never change the sign of the trace in any block. But this is
required for obtaining the inverse entries.

We show that the number of time steps cannot be
constant, i.\,e. independent of $n$. Assume w.l.o.g. that
$tr(A)>0$.  Choose a partition of
$SO(3)=S_1\cup S_2\cup\ldots\cup S_p$  into
equivalence classes such that for any pair $O,\tilde{O}\in S_i$
($i=1,\ldots,p$) implies $tr(O A \tilde{O}^T)\geq 0$ ($p$ depends on $A$). 
This is always possible since the partition can be chosen in such
a way that all orthogonal transformations in the same class
are sufficiently close together.
Then
$O A \tilde{O}^T$ almost corresponds to a conjugation of $A$ by an
orthogonal matrix and its trace cannot deviate too much from $tr(A)$
due to the continuity of the trace.

Let $O_j^{(k)}\in SO(3)$ 
be the operation performed on the $k$th spin in time step
$j$ (described as action on the coupling matrix). In each time step at
least $\lceil n/p\rceil$ spins are transformed by operations of the
same class\footnote{Here $\lceil x\rceil$ denotes the smallest integer
$m$ such that $x\le m$.}. After $N$ time steps at least $\lceil
n/p^N\rceil$ spins have been transformed by operations of the same
class in each step.  Let $k$ and $l$ be such a pair that has been
transformed by operations of the same equivalence class in each time
step. Then the trace of the average coupling between the spins $k$ and
$l$ after the sequence is given by
\begin{equation}\label{tracepositive}
tr(\sum_j t_j O_j^{(k)} A {O_j^{(l)}}^T)\ge 0\,.
\end{equation}
If the sequence is an inversion scheme then we must have
\[
\sum_j t_j O_j^{(k)} A {O_j^{(l)}}^T = -A\,.
\] 
But this condition contradicts the
inequality~(\ref{tracepositive}). Therefore such a pair $k$ and $l$
must not exist. This gives the lower bound $N\ge\log n/ \log p$ on 
the number of time steps $N$.

We show that the time overhead does not depend on $n$. We assume
w.l.o.g. $a_x>0>a_z$ for the eigenvalues of $A$. First we describe a
partial decoupling scheme selecting for instance the $\sigma_z\otimes
\sigma_z$ coupling terms while switching off the $xx$ and $yy$
terms. Following \cite{reval} such a decoupling can be achieved by
certain sequences of local conjugations by the unitary
$i\sigma_z$. Each time step of this scheme is described by a column of
a Hadamard matrix. The entries determine which spins are conjugated by
$i\sigma_z$ transformations. The idea is that the $xx$ and $yy$ terms
acquire in exactly half of the time steps a minus sign. Note that this
scheme does not weaken the $zz$ coupling. In other words, although
this scheme requires a number of {\it time steps} of the order $n$ the
{\it time overhead} for the simulation of the interaction $\sum w_{kl}
a_z \sigma_z^{(k)}\sigma_z^{(l)}$ by the original one is $1$.

There are two cases depending on whether $a_y\ge 0$ or $a_y<0$.
Consider the case $a_y\ge 0$ w.l.o.g. First switch off the $xx$ and
$yy$ interactions. Apply on each spin a conjugation by $u_y$ (defined
as in the beginning). This simulates the interaction
\[
\sum_{k<l} w_{kl}\, a_z\, \sigma_x^{(k)}\sigma_x^{(l)}\,.
\]
By applying this interaction for the (relative) time $-a_x/a_z$ we
have reversed the $xx$-components in the original Hamiltonian. Since
$a_y$ is not negative we can reverse similarly the $yy$-components
with time overhead $-a_y/a_z$. To invert the $zz$-components we use
the $xx$-components. The time overhead is $-a_z/a_x$. In summary we
see that the time overhead is independent of $n$. It only depends on
the eigenvalues of $A$.

\vspace{0.3cm}
{\bf Case 3.} Let $t_1,t_2,\dots,t_N$ be the (relative) times and 
$V_1,V_2,\dots,V_N$  with
$V_j\in \bigoplus_{k=1}^n SO(3)$  be the operations of an inversion 
scheme, i.\,e.,
\begin{equation}\label{Anf}
\sum_j t_j V_j (W\otimes A)  V_j^T = - W\otimes A\,.
\end{equation}

First note that an inversion scheme for a given Hamiltonian $H_J$ can
also be applied to a rescaled Hamiltonian obtained by changing the
strength and the sign of an arbitrary spin pair interaction. If all
the coefficients $w_{kl}$ are non-vanishing for $k\neq l$ then we can
equivalently describe a time inversions scheme for the coupling matrix
$J:=\tilde{W} \otimes A$ where all non-diagonal entries of $\tilde{W}$
are $1$ and the diagonal consist of zeros. Hence the fact that
interactions decrease with distance of the spins is irrelevant for
considerations of the complexity of time inversion (as long as the
interaction cannot be neglected). Therefore we assume w.l.o.g. all
non-diagonal entries of $W$ to be $1$.

For each time step let $V_j$ be the block diagonal matrix
$V_j=U_j^{(1)}\oplus\cdots\oplus U_j^{(n)}$. Set $R:=\sum_{j=1}^N t_j
V_j ({\mathbf 1}\otimes A) V_j^T$. We add $R$ to both sides of
eq.~(\ref{Anf}) and obtain
\begin{equation}\label{KA2}
\sum_{j=1}^N t_j V_j \big((W+ {\mathbf 1})\otimes A\big) V_j^T = 
-W\otimes A\, + R \,.
\end{equation}
The rank of the matrix $(W+{\mathbf 1})$ is $1$ since all its entries
are $1$.  Consequently, the rank of the left hand side of equation
(\ref{KA2}) is at most $N\,r$, where $r$ is the rank of $A$.  Note
that $R$ is a semi-positive matrix. Therefore the rank of the right
hand side is at least the number of positive eigenvalues of $-W\otimes
A$. The matrix $-W$ has $n-1$ positive eigenvalues (by adding the
identity to $W$ one easily verifies that $W$ has the eigenvalues
$n-1,-1,-1,\dots,-1$).  Hence $-W\otimes A$ has $(n-1) r$ positive
eigenvalues. We conclude that the rank of the right hand side is at
least $(n-1)r$. This proves the statement $N\geq n-1$.

We show now that the time overhead also grows with $n$. Let
$\lambda_{\min}$ and $\lambda_{\max}$ be the minimal and maximal
eigenvalues of $J$, respectively. Note that $\lambda_{\min}$ is
negative since $J$ is traceless. Let $t_1,t_2,\dots,t_N$ be the
(relative) lengths of the time steps. Then we have
\[
\sum_j t_j \lambda_{\max} \leq -\lambda_{\min},
\]
since the maximal eigenvalue of the sum of matrices in
eq.~(\ref{condition}) is at most the sum of their maximal eigenvalues
\cite{bha} and $-\lambda_{\min}$ is the maximal eigenvalue of $-J$.
We conclude that the time overhead $\tau:=\sum_j t_j$ for the inverse
evolution is at least
\[
\tau\geq -\lambda_{\max}/\lambda_{\min}\,.
\]
Since the maximal and minimal eigenvalues of $W$ are $(n-1)$ and $-1$,
respectively, the maximal and minimal eigenvalues of $J$ are
$(n-1)a_{\max}$ and $-a_{\max}$ where $a_{\max}$ is the maximal
eigenvalue of $A$. We conclude that the minimal time overhead is at
least $n-1$.$\Box$
\vspace{0.3cm}

The truncated dipole-dipole coupling is an example for case 1.
The corresponding matrix $A$ is given by $A:=diag \, (1,1,-2)$.
An important example for case 3 is the strong scalar coupling where
$A$ is the identity matrix. By combining both types one can easily find
examples for case 2 \cite{Gl}.

Besides the practical applications of refocusing the lower bounds on
the inversion complexity may be seen as a small contribution to the
old debate on the origin of phenomenological irreversibility: Whereas
the considered dynamical evolutions are strictly reversible the
complexity of a real implementation of the time inversion is growing
with the number of interacting particles.  Note that the relevance of
complexity theory for the definition of physical entropy has been
advocated for by several authors \cite{zurek,zurek2,li}. Of course our
complexity bounds are restricted to the assumptions of average
Hamiltonian theory. Inversion schemes that are not based on this
approach and include higher order terms of the Magnus expansion may be
less complex. However, our results present an example how to develop a
complexity theory of time reversal in physics.

%%%%%%%%%%%%%%%%%%%%%%%%%%%%%%%%%%%%%%%%%%%%%%%%%%%%%%%%%%%%
%
% The Literature
%
%%%%%%%%%%%%%%%%%%%%%%%%%%%%%%%%%%%%%%%%%%%%%%%%%%%%%%%%%%%%

%\nocite{*}

\end{multicols}

\end{document}